\documentclass[trackchanges,twocolumn]{aastex701}

\usepackage{multirow}
\usepackage[T1]{fontenc}
\usepackage[utf8]{inputenc}
\usepackage{soul}
\usepackage{xcolor}
\sethlcolor{yellow}

\begin{document}

\title{Convolutional Neural Network for Extraction of $n = 1$ Photon Ring of Black Holes}

\author[orcid=0009-0002-2194-4898]{Courtney Duong}
\affiliation{Department of Physics, University of California, Santa Barbara, CA 93106-9530, USA}
\email[show]{courtneyduong@ucsb.edu}  
\author[orcid=0009-0008-8192-928X]{Frank Myhre}
\affiliation{Department of Physics, University of California, Santa Barbara, CA 93106-9530, USA}
\email{fmyhre@ucsb.edu}

\author[0000-0003-4914-5625]{Joseph R. Farah}
\affiliation{Miller Institute for Basic Research in Science, 206B Stanley Hall, Berkeley, CA 94720, USA}
\affiliation{Department of Astronomy, University of California, Berkeley, CA 94720-3411, USA}
\email{josephfarah@berkeley.edu}

\begin{abstract}

Very-long-baseline interferometry (VLBI) may enable direct imaging of fine, event-horizon scale black hole structures such as the photon ring. In particular, the $n=1$ photon subring encodes information about black hole spin in its radial profile and azimuthal brightness modulation, making it a key proxy for spacetime properties. However, the $n=1$ subring overlaps with emission from the $n=0$ subring in observations, requiring isolation of the $n=1$ subring for measurements. We present a convolutional neural network (CNN) capable of extracting the $n=1$ subring from composite images containing $n=0$ and $n=1$ subrings. We simulate $111,000$ $m$-ring images with \texttt{eht-imaging} to train and validate the CNN. The CNN accurately recovers the radial and angular brightness profiles of the $n=1$ subring with a mean normalized cross correlation (NXcorr) of $\sim0.99$. The CNN provides a detection capability by predicting no false positives when input images lack a $n=1$ subring. We perform feature extraction with \texttt{ringfit} to show that the CNN-predicted $n=1$ subring accurately recovers the ground truth radial profiles and azimuthal brightness modulation, demonstrating the CNN's capability to extract spin-sensitive features. We further test the CNN on black hole images generated by \texttt{KerrBAM} and general relativistic magnetohydrodynamic (GRMHD) simulations, finding that it recovers the overall $n=1$ subring radial profiles while exhibiting discrepancies in the recovered intensity profiles. These results demonstrate the potential of deep learning methods to isolate the $n=1$ subring from overlapping emission, providing a framework for analyzing future high resolution black hole images from proposed VLBI missions such as the Black Hole Explorer (BHEX).

\end{abstract}

%% https://astrothesaurus.org

\section{Introduction} 

Very-long-baseline interferometry (VLBI) has enabled direct imaging of black holes at unprecedented angular resolutions, providing access to structures near the event horizon that were previously unresolved in sources such as M87* and Sgr A* \citep{EHT_Collab_2022}. Continued improvements in angular resolution and baseline coverage may make it possible to resolve increasingly fine structures formed near the event horizon, such as the photon ring. Several proposed and developing VLBI efforts aim to extend black hole imaging to these smaller angular scales and higher sensitivities \citep{pesce2019,akiyama2026,trippe2026}. One such effort is the Black Hole Explorer (BHEX) mission, which would extend the Event Horizon Telescope (EHT) to space and improve the angular resolution of direct images from 20 $\mu$as to as fine as 6 $\mu$as \citep{Akiyama2024BHEXJapan, Johnson_2024}. This improvement in angular resolution will provide access to finer event-horizon scale features beyond the coarse shadow features currently resolved by the EHT.

The \textit{photon ring} is a structure around a black hole predicted from general relativity. A photon ring arises from a shell of unstable photon orbits and consists of $n$ subrings produced by photons completing $n$ half-orbits around a black hole \citep{Luminet1979, Johnson_2020,GrallaLupsascaLensing}. The $n = 0$ subring is mainly shaped by fluid flow while higher-order ($n \geq 1$) subrings are primarily dependent on the black hole spacetime properties such as spin. Recent literature show that specifically the $n = 1$ subring encodes information about black hole spin in its radial and angular brightness profiles, which can be approximately modeled by a shape known as an $m$-ring \citep{Johnson_2020}. Spin can potentially be extracted from the $n = 1$ direct image with current methods \citep{Broderick_2022, palumbo2022, Paugnat_2022, chang2024, farah_2024, farah2025, keeble2025}. The projected angular resolution of BHEX is expected to be sufficient to resolve the $n = 1$ subring for M87* and Sgr A* \citep{Johnson_2024}, making BHEX a useful framework for studying the image analysis challenges associated with extracting this structure from future high resolution black hole images.

These images will be contaminated with the $n = 0$ subring, with negligible contributions from higher-order ($n \geq 2$) subrings. The key challenge then is to isolate the $n = 1$ subring from the $n = 0$ subring.

In this paper, we present a deep-learned convolutional neural network (CNN) to isolate the $n = 1$ photon subring from dominant $n = 0$ emission. The CNN is trained on simulated $m$-ring images containing combined $n = 0$ and $n = 1$ subrings in order to extract the $n = 1$ subring and recover its spin-sensitive observables, such as size and azimuthal brightness modulation. We evaluate the CNN’s performance on reconstruction accuracy using the feature extraction software \texttt{ringfit}. Our results demonstrate robust proof-of-concept extraction of the $n = 1$ subring, laying the groundwork for an accurate method to probe spin-sensitive observables from future BHEX observations.

\section{Methods} \label{sec:style}

\subsection{Generation of training data} \label{sec:dataset_section}

\begin{figure*}[ht]
    \centering
    \includegraphics[width=0.6\textwidth]{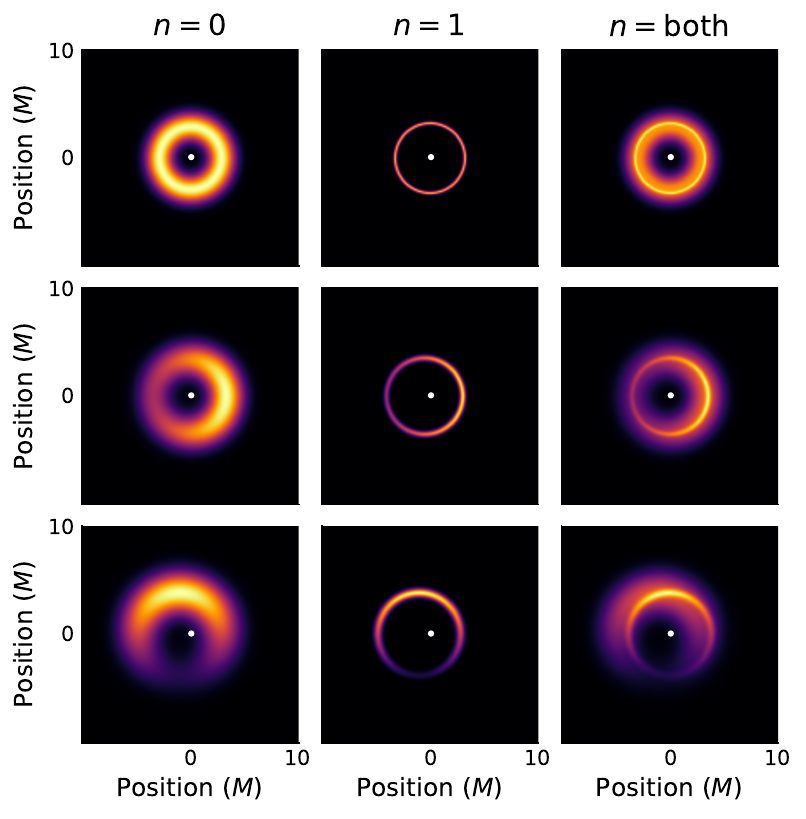}
    \caption{Sample photon subrings from dataset of $\sim111,000$ generated images. Each column shows individual and composite subring images: (left) $n=0$ subring images, (center) $n=1$ subring images, and (right) $n=\text{both}$ images containing both subrings. Each row shows a different parameter configuration: (top, center, bottom) $F_0=$ (0.5, 0.65, 0.8) Jy, $d_0=$ (7.9, 9.2, 10.5) M, $\alpha_0$ = (1.3, 1.6, 2.0) M, $\Delta x_0$ = (0, 0.3, 0.7) M, $\beta_{s,0}$ = (0, 0.2, 0.4); $\theta_0$ = (0, 90, 180)$^\circ$, $F_1$ = (0.02, 0.51, 1.0) Jy, $d_1$ = (8.4, 9.2, 10.0) M, $\alpha_1$ = (0.1, 0.3, 0.4) M, $\Delta x_1$ = (0, 0.3, 0.7) M, $\beta_{s,1}$ = (0, 0.2, 0.4), $\theta_1$ = (0, 90, 180)$^\circ$.}
    \label{fig:mring_dataset_construction}
\end{figure*}

To train and validate the CNN, we generate over 111,000 synthetic images with multiple subrings using the m-ring model \citep{Johnson_2020} from the \texttt{eht-imaging} \citep{eht-imaging} Python package. The generated dataset consists of intensity-only images with an image resolution of 128x128 pixels, normalized to values between 0 to 1. 

We show all images in units of gravitational radius $(0,M)$ (Figure \ref{fig:mring_dataset_construction}), allowing comparison across black holes independent of their actual mass or distance. Since increasing mass enlarges the black hole while increasing distance reduces its apparent size, only the mass-to-distance ratio determines the observed angular size. Our analysis is therefore insensitive to the specific mass of the black hole, as long as the corresponding mass-to-distance ratio and angular resolution allows the shadow to be resolved. When a conversion to angular units (e.g. microarcseconds) is relevant, we assume a mass-to-distance ratio estimated for M87* from EHT observations. 

The training dataset also includes images without an $n=1$ subring (i.e. $n=0$ subring images), enabling the CNN to function as a detector and reducing its tendency to ‘hallucinate’ an $n=1$ subring (i.e., generate a feature that is not present in the input).

The training dataset features significant parameterization flexibility for both subrings including direction and magnitude of brightness asymmetry, sizes, and positional offsets. We choose and implement a range of parameter values that extend beyond the physically expected ranges of realistic black hole simulations, consequently exposing the CNN to a broader range of subring configurations. For physically realistic Kerr black holes, photon ring size and asymmetry are constrained by black hole spin, inclination, and emission geometry \citep{Farah_2020_BHShadowCurve, farah_2024}. Kerr black hole models with emission radii of $3M-7M$ show that the size and morphology of the $n=1$ subring vary with black hole spin and inclination, with the smallest enclosing ellipse (SEE) capturing additional variation in the subring size and brightness distribution \citep{farah_2024}. In comparison, our dataset allows $n=0$ subring diameters of $7.0-10.5M$ and $n=1$ subring diameters of $8.4-10.0M$, along with brightness asymmetry strengths $\beta_{s,0}$ and $\beta_{s,1}$ ranging from $0$ to $0.4$. Thus, our subring parameter space allows larger variations in subring size and brightness asymmetry than those expected for physically realistic black hole configurations. This expanded parameter space is intended to build robustness for application to realistic simulations, which occupies a narrow subset of our parameter space. Parameter values for the subrings used in CNN training are summarized in Table \ref{tab:subring_parameters}. For both the $n=0$ and $n=1$ subrings, we vary flux ($F_0$, $F_1$), diameter ($d_0$, $d_1$), width ($\alpha_0$, $\alpha_1$), horizontal offset ($\Delta x_0$, $\Delta x_0$), asymmetry strength ($\beta_{s,0}$, $\beta_{s,1}$), and asymmetry angle ($\theta_0$, $\theta_1$).

In the original dataset, we generated $\sim100$ $n=0$ subring images and $\sim700$ $n=1$ subring images, each with a different parameter configuration, totaling to $\sim70,000$ composite images. We then augmented the dataset by $41,000$ images by randomly selecting $1000$ different parameter configurations where both subrings are aligned and rotating the $n=1$ subring relative to the $n=0$ subring by $41$ different angles, $0^\circ-30^\circ$ in $1^\circ$ steps to cover small brightness asymmetry angles and $45^\circ-180^\circ$ in $15^\circ$ steps to cover large brightness asymmetry angles.

\begin{table*}[!t]
\centering
\begin{tabular}{llcc}
\hline
Subring & Parameter & Original simulated values  \\
\hline
\multirow{6}{*}{$n=0$}
& $F_0$ & 0.5, 0.65, 0.8 Jy \\
& $d_0$ & 7.9, 9.2, 10.5 M \\
& $\alpha_0$ & 1.3, 1.6, 2.0 M \\
& $\Delta x_0$ & 0, 0.3, 0.7 M \\
& $\beta_{s,0}$ & 0, 0.2, 0.4 \\
& $\theta_0$ & 0, 90, 180$^\circ$ \\
\hline
\multirow{6}{*}{$n=1$}
& $F_1$ & 0.02, 0.51, 1.0 Jy \\
& $d_1$ & 8.4, 9.2, 10.0 M \\
& $\alpha_1$ & 0.1, 0.3, 0.4 M \\
& $\Delta x_1$ & 0, 0.3, 0.7 M \\
& $\beta_{s,1}$ & 0, 0.2, 0.4 \\
& $\theta_1$ & 0, 90, 180$^\circ$ \\
\hline
\multicolumn{3}{l}{\textbf{Augmentation}} \\
\hline
\multicolumn{2}{l}{Selected configurations} & 1,000 \\
\multicolumn{2}{l}{Transformation} & $n=1$ rotated relative to $n=0$ \\
\multicolumn{2}{l}{Rotation angles} & $0^\circ-30^\circ$ ($1^\circ$ steps), $45^\circ-180^\circ$ ($15^\circ$ steps) \\
\hline
\end{tabular}
\caption{Subring parameter values for the original simulated dataset and the augmentation applied to a subset of the simulated images. The listed values indicate the values represented in the dataset; not all possible combinations of these values were generated.}
\label{tab:subring_parameters}
\end{table*}

A sample of the dataset is shown in Figure \ref{fig:mring_dataset_construction}. The images mimic characteristic features of photon rings, where each successive subring is exponentially narrower and dimmer than the preceding one \citep{Johnson_2020}.

\subsection{Construction of CNN}

\begin{figure*}[t]
    \centering
    \includegraphics[width=\textwidth]{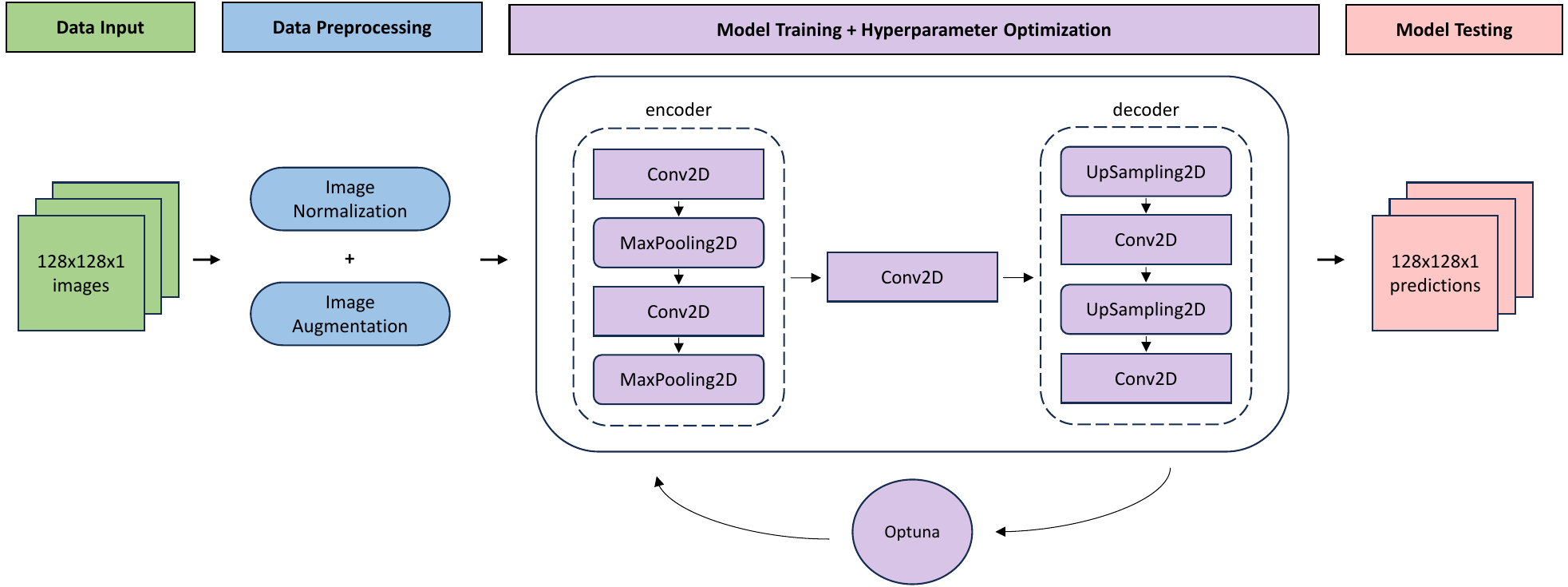}
    \caption{CNN framework and workflow. The network takes the augmented dataset of normalized 128x128x1 images as input. The encoder-decoder architecture, comprising convolutional, max pooling, and upsampling layers, is trained to reconstruct the target output, producing predicted images of input dimensions during testing. Training occurs over 100 trials, with \texttt{Optuna} evaluating a different hyperparameter combination in each trial by minimizing the validation loss corresponding to the selected loss function. After the hyperparameter optimzization is complete, predictions are generated with the configuration selected by the Optuna search and subsequently evaluated.}
    \label{fig:cnn layout figure}
\end{figure*}

CNNs are multilayered, gradient-based learning algorithms, suitable for image recognition and classification tasks \citep{lecun_1998, Krizhevsky_2012}. CNNs learn hierarchical spatial features directly from raw pixel intensities, enabling them to detect progressively complex patterns. However, training CNNs on high-resolution datasets can require substantial computational time and memory due to the large number of input dimensions and parameters.

Previous work has demonstrated the use of deep learning for black hole parameter recovery from simulated images (e.g. \citealt{vanderGucht2020,farah_2024}). We develop a similar convolutional framework to extract the $n=1$ photon ring from simulated images.

We implement a CNN for image-to-image translation using the TensorFlow 2.19 Keras API \citep{chollet2015}. The network follows an encoder-decoder structure \citep{hinton2006} shown in Figure \ref{fig:cnn layout figure}, in which the encoder compresses the input into a compact representation and the decoder reconstructs it back to the original spatial dimensions. The encoder consists of two convolutional layers with activation functions and ‘same’ padding, each followed by a max pooling layer \citep{ranzato2007} to progressively reduce spatial dimensions while increasing the number of learned feature maps. A bottleneck layer further transforms these feature maps into a compact representation before reconstruction by the decoder. The decoder mirrors this structure by applying upsampling and convolutional layers to restore the spatial dimensions of the input image. The final output layer applies a 1x1 convolution with a sigmoid activation function to produce an image of the same size as the input image. This architecture is well-suited to the present task because the encoder enables the CNN to learn increasingly abstract features such as intensity variations and subring edges in the input image, while the decoder uses these features to reconstruct the spatial location and structure of the $n=1$ subring.

\begin{table*}[!t]
\centering
\begin{tabular}{llcc}
\hline
Hyperparameter & Search space \\
\hline
Learning rate & $10^{-6}-10^{-3}$, log scale \\
Batch size & 8, 16, 32, 64 \\
Loss function & Binary cross-entropy, MSE, MAE, Sobel, SSIM \\
Optimizer & Adam, AdamW \\
Number of filters, layer 1 & 32, 64 \\
Number of filters, layer 2 & 64, 128 \\
Number of filters, layer 3 & 128, 256 \\
Kernel size & 3x3, 5x5 \\
Activation function & ReLU, Leaky ReLU, GELU \\
\hline
\end{tabular}
\caption{CNN hyperparameter search space explored using Optuna.}
\label{tab:model_hyperparameters}
\end{table*}

We fixed the overall architecture while optimizing hyperparameters such as learning rate, batch size, loss function, optimizer, number of filters, kernel size, and activation function using \texttt{Optuna} \citep{akiba2019}, a hyperparameter optimization framework. We evaluated different hyperparameter combinations summarized in Table \ref{tab:model_hyperparameters} over 100 trials. The number of training epochs was fixed at 50 for all trials. The hyperparameter configuration selected by the Optuna search uses a mean squared error (MSE) loss function, the AdamW optimizer \citep{loshchilov2019}, a learning rate of $1.94\times10^{-4}$, a batch size of 16, a filter configuration of 32-128-128, a kernel size of $3$x$3$, and a Leaky Rectified Linear Unit (Leaky ReLU) activation function \citep{hinton2010, Maas2013}. This configuration was selected by minimizing the validation loss corresponding to the validation loss function used in each trial.

\subsection{Training and validating the CNN} \label{training_validating}

We allocated 80\% of the full dataset for training and 20\% for validation.

Training follows the gradient-based learning framework described by \cite{lecun_1998}, in which network parameters (weights and biases) are iteratively updated over multiple epochs to minimize a loss function (e.g. MSE). The discrepancy between CNN predictions and ground truth is quantified by the loss function as images are passed through the network. Nonlinear activation functions (e.g. Leaky ReLU) at each layer enables the network to learn complex features. 

The validation subset is not seen during training to prevent overfitting, ensuring that performance metrics reflect the network’s ability to generalize and reconstruct unseen data. We compared the CNN’s predictions to corresponding ground truth images using normalized cross correlation (NXcorr). We use this metric following the definition and application in \cite{farah2022} where NXcorr is used as a measure to identify portions of reconstructions that most closely match a model. The NXcorr $\rho_{NX}(X,Y)$ between two images $X$ and $Y$ is defined as 
\begin{equation}
    \label{eq:nxcorr}
    \rho_{NX}(X,Y)=\frac{1}{N}\sum_{i}\frac{(X_i-\langle X \rangle)(Y_i-\langle Y \rangle)}{\sigma_X\sigma_Y}
    \end{equation}
\citep{eht2019}. NXcorr quantifies the similarity between images by measuring the correlation between pixel intensity distributions. An NXcorr value of 1 indicates perfect structural agreement between images, while a value of 0 corresponds to little to no structural similarity.

We use NXcorr to evaluate CNN performance by comparing the similarity of the CNN predictions to the ground truth $n=1$ subring images relative to the similarity of the input images containing both $n=0$ and $n=1$ subrings to the same ground truth. This comparison quantifies how accurately the CNN recovers the $n=1$ subring.

Prior to calculating NXcorr values between the CNN’s predictions and ground truth images, we filtered out samples without an $n=1$ subring. In such cases, all pixels in the true $n=1$ subring image are zero, so any tiny noise-like deviations in the predicted image would yield low similarity values that do not reflect actual CNN performance. We also excluded images where $n=1$ subring alone dominated the image and the $n=0$ subring component was comparatively faint. Such cases are relatively easy for the CNN to reconstruct the $n=1$ subring since it is already the most prominent feature in the image, so the resulting NXcorr is not particularly informative as it is already high between the $n=1$ subring and combined subring images. The radial profiles of the subrings are approximately Gaussian and for a Gaussian profile, the peak intensity increases as the profile becomes narrower for a fixed flux. The $n=0$ subring is approximately 10 times broader than the $n=1$ subring, so the $n=1$ subring can have a substantially higher peak intensity even when its flux is lower than that of the $n=0$ subring. Therefore, flux alone does not determine the prominence of the two subrings in the image. To meaningfully assess the network’s performance, we focused on cases where the $n=1$ subring signal must be recovered in the presence of a bright $n=0$ subring, hence providing a thorough test of the CNN’s ability to disentangle overlapping subring flux contributions. 

\begin{figure}[h]
    \centering
    \includegraphics[width=0.45\textwidth]{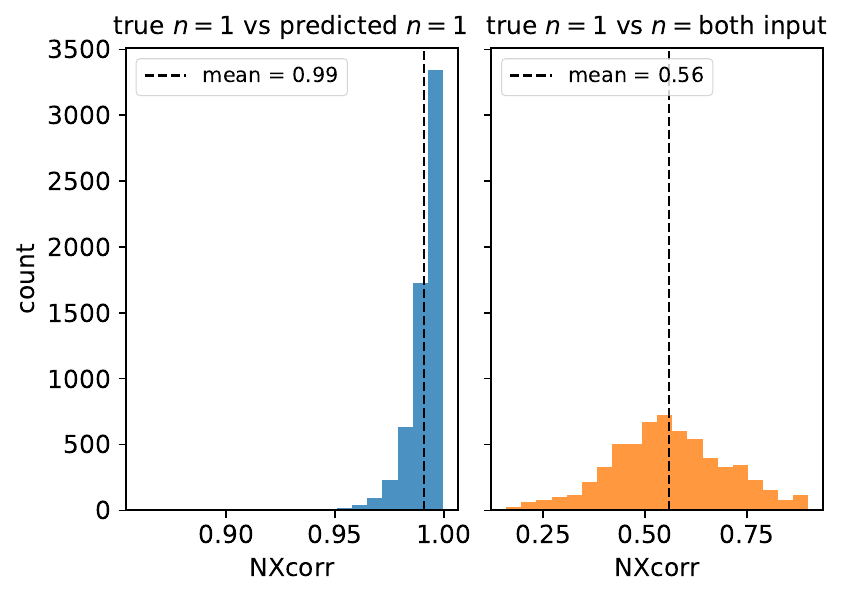}
    \caption{Histograms of normalized cross correlation (NXcorr) distributions comparing similarity to the ground truth $n=1$ subring images, described in Section \ref{training_validating}.}
    \label{fig:nxcorr}
\end{figure}

Figure \ref{fig:nxcorr} shows histograms of the resulting NXcorr distributions. In both panels, the x-axis denotes the NXcorr values and the y-axis indicates the number of images at each similarity level. The left panel compares the CNN predictions with the ground truth, showing a sharply peaked distribution with a mean NXcorr of 0.99. This indicates consistently strong similarity between CNN predictions and ground truth, hence that the CNN accurately reconstructs the $n=1$ subring. The right panel compares the input images with the ground truth, producing a broader distribution centered at a lower mean NXcorr of 0.56, reflecting weaker similarity which is consistent with the expectation that the CNN input does not correspond to the true $n=1$ subring structure.

\section{Results} \label{sec:floats}

\begin{figure*}[t]
    \centering
    \includegraphics[width=0.7\textwidth]{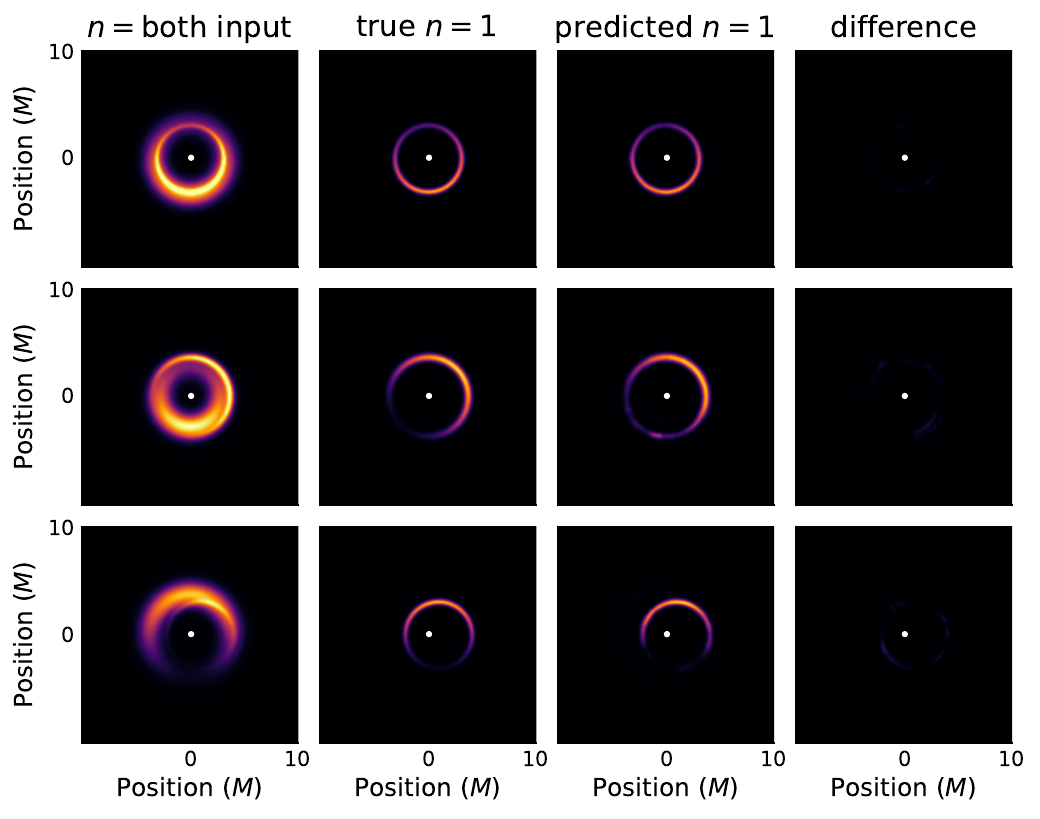}
    \caption{Sample photon subring images used to evaluate CNN performance. (Left column) input images with all subrings, (center left column) ground truth $n=1$ subrings, (center right column) CNN predicted $n=1$ subrings, and (right column) corresponding difference images. The ground truth and CNN predicted $n=1$ subring images are normalized to the dimmest and brightest pixel values in the input images. Each row shows a different parameter configuration: (top, center, bottom) $F_0=$ (0.65, 0.65, 0.5) Jy, $d_0=$ (9.2, 7.9, 10.5) M, $\alpha_0$ = (1.3, 1.3, 1.3) M, $\Delta x_0$ = (0, 0, 0) M, $\beta_{s,0}$ = (0.2, 0.2, 0.4); $\theta_0$ = (0, 0, 180)$^\circ$, $F_1$ = (0.02, 0.02, 0.02) Jy, $d_1$ = (8.4, 10.0, 8.4) M, $\alpha_1$ = (0.3, 0.4, 0.3) M, $\Delta x_1$ = (0, 0, 0.7) M, $\beta_{s,1}$ = (0.2, 0.4, 0.4), $\theta_1$ = (16, 135, 180)$^\circ$.} The close correspondence between the ground truth and predicted subring images demonstrate the CNN's ability to recover the $n=1$ subring structure from composite inputs.
    \label{fig:mring_all_rings_with_diff}
\end{figure*}

Our CNN, trained on the dataset described in Section \ref{sec:dataset_section}, effectively extracts the $n=1$ subring image from the combined subring input images. The extracted images accurately recover the radial and angular brightness profiles of the ground truth $n=1$ subring image, demonstrating that the CNN not only identifies the $n=1$ subring structure but also preserves physical features of the photon ring required for spin measurement. 

Comparisons between the input images, the ground truth $n=1$ images, and the CNN-predicted $n=1$ images are shown in Figure \ref{fig:mring_all_rings_with_diff}. The difference images, computed as subtraction of the predicted $n=1$ subring from the ground truth $n=1$ image, highlight only minor regions of intensity deviations, demonstrating that the CNN recovers the dominant radial and angular brightness features of the $n=1$ subring. Notably, the CNN extracts the azimuthal brightness modulation of the $n=1$ subring correctly despite the differing brightness asymmetry of the other subring in the composite input image, demonstrating the CNN's ability to accurately probe the $n=1$ subring. 

\begin{figure}[h]
    \centering
    \includegraphics[width=0.45\textwidth]{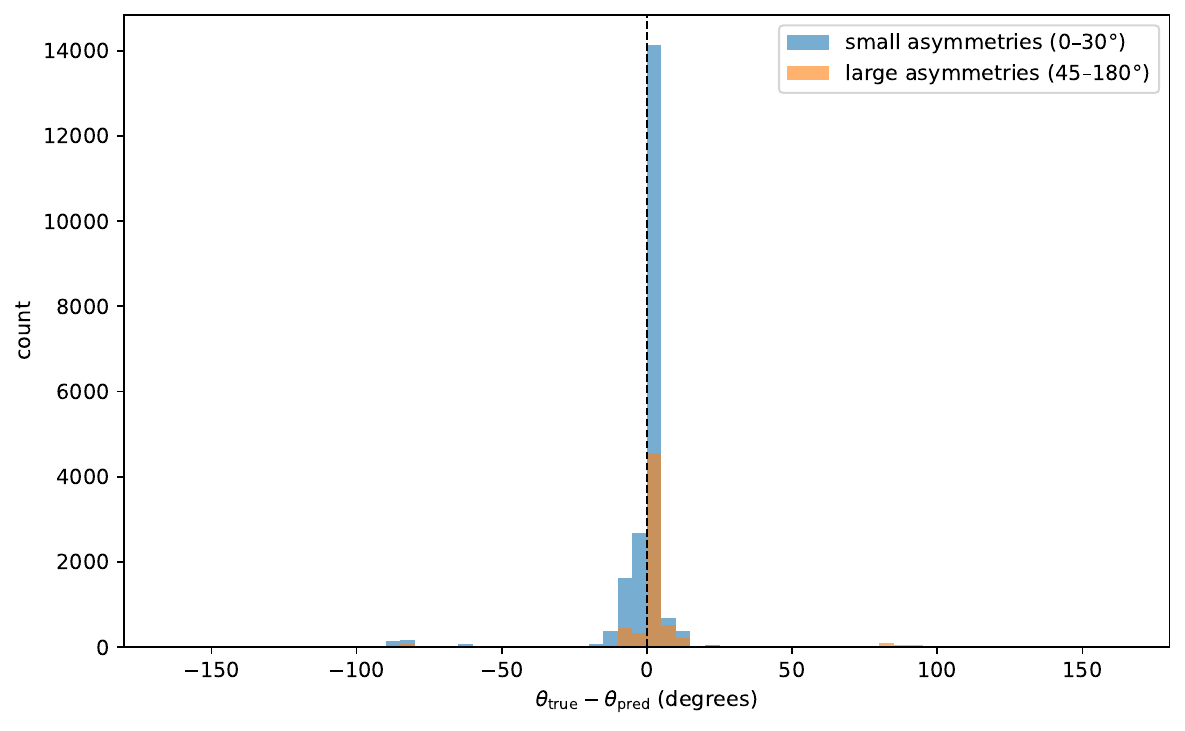}
    \caption{Histograms of brightness asymmetry angle recovery at small asymmetries ($0-30^\circ$, $1^\circ$ increments) and large asymmetries ($45-180^\circ$, $15^\circ$ increments), described in Section \ref{sec:floats}.}
    \label{fig:brightness_angle_histogram}
\end{figure}

To quantify the CNN's ability to recover brightness asymmetry at both small ($0-30^\circ$, $1^\circ$ increments) and large angles ($45-180^\circ$, $15^\circ$ increments), we consider subring images where $\beta_{s,1} \not= 0$ and for each, calculate the difference between the true $n=1$ peak intensity angle and the predicted $n=1$ peak intensity angle, $\theta_{1,\text{true}}-\theta_{1,\text{pred}}$. As shown in Figure \ref{fig:brightness_angle_histogram}, we find that $93\%$ of brightness asymmetry angle recovery is clustered around $\theta_{1,\text{true}}-\theta_{1,\text{pred}}=0^\circ$ with a spread of $\sigma_s=4.1^\circ$ for small brightness asymmetry angles and a spread of $\sigma_l=4.6^\circ$ for large brightness asymmetry angles. These results demonstrate that the CNN accurately recovers brightness asymmetry angle across both small and large angular offsets.

\begin{figure*}[ht]
    \centering
    \includegraphics[width=0.7\textwidth]{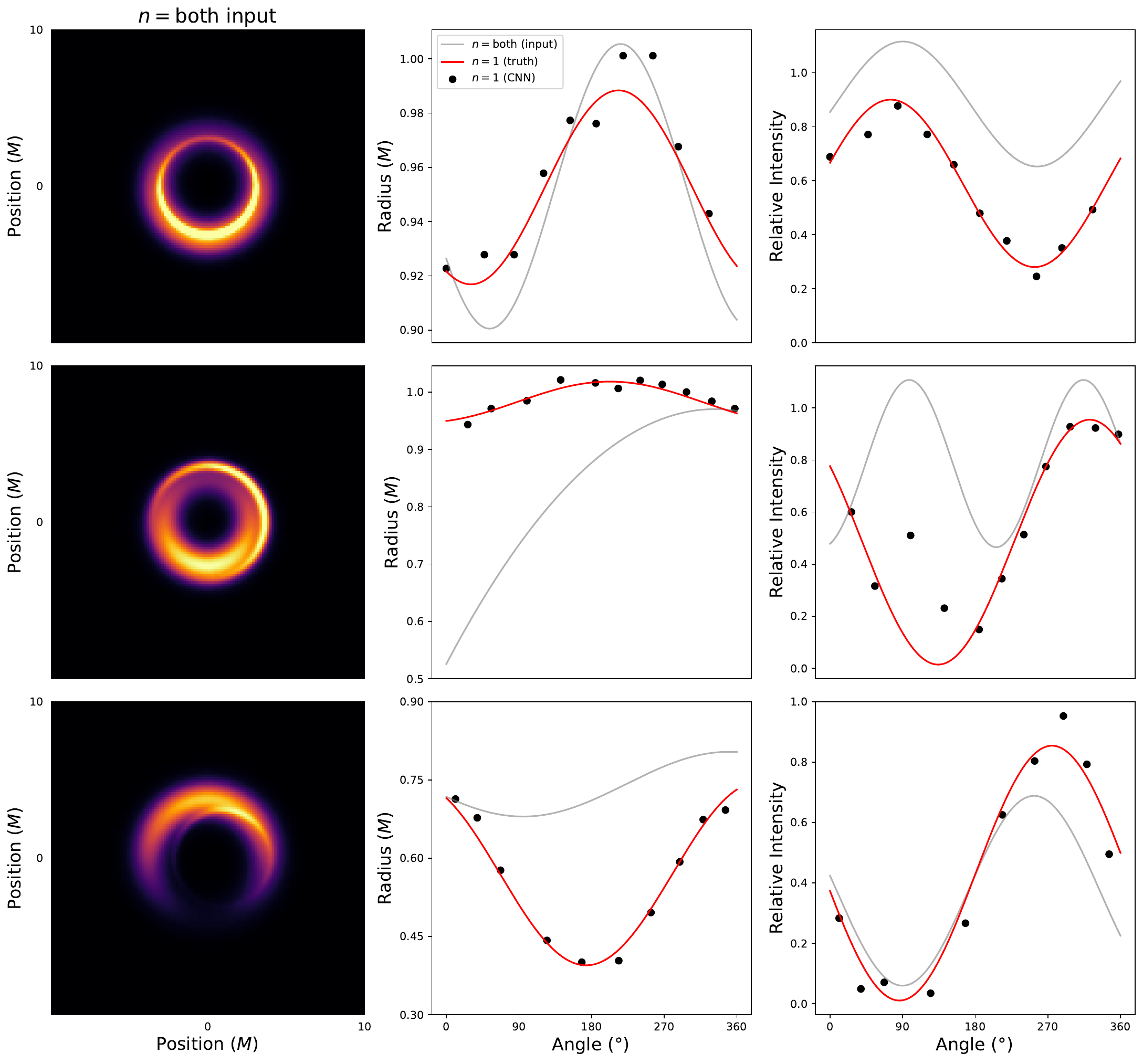}
    \caption{Extraction of radial and angular intensity profiles by \texttt{ringfit} performed on sample photon subring images in Figure \ref{fig:mring_all_rings_with_diff}. (Left) input images with all subrings; (center) extracted radial profiles for each image's input image, ground truth $n=1$ subring, and CNN predicted $n=1$ subring; and (right) extracted intensity profiles for the same image components. The ground truth radial and azimuthal intensity functions are sinusoidally interpolated (solid lines) for comparison to the CNN extraction (points). The CNN accurately recovers the radial and angularbrightness profiles of the $n=1$ subring as the radial and angular brightness extractions for CNN predicted $n=1$ subring (black) closely corresponds with the ground truth $n=1$ subring extractions (red), rather than the input image extractions (gray).}
    \label{fig:feature_extraction}
\end{figure*}

We apply a feature extraction software, \texttt{ringfit} \citep{ringfit}, to map the radial and angular brightness profiles of our photon ring images (Figure \ref{fig:feature_extraction}). The radial and angular brightness profiles of the predicted $n=1$ subrings closely follow that of the ground truth $n=1$ subrings, rather than the composite inputs, indicating that the CNN successfully recovers the geometric and intensity structures of the $n=1$ subring. However, as shown in the first sample in Figure \ref{fig:feature_extraction}, the predicted $n=1$ subring shows a near-uniform intensity offset relative to the ground truth. The CNN correctly recovered the brightness asymmetry pattern, but the lower recovered intensity indicates that the CNN can introduce a small but systematic bias across the ring.

\begin{figure*}[h]
    \centering
    \includegraphics[width=0.7\textwidth]{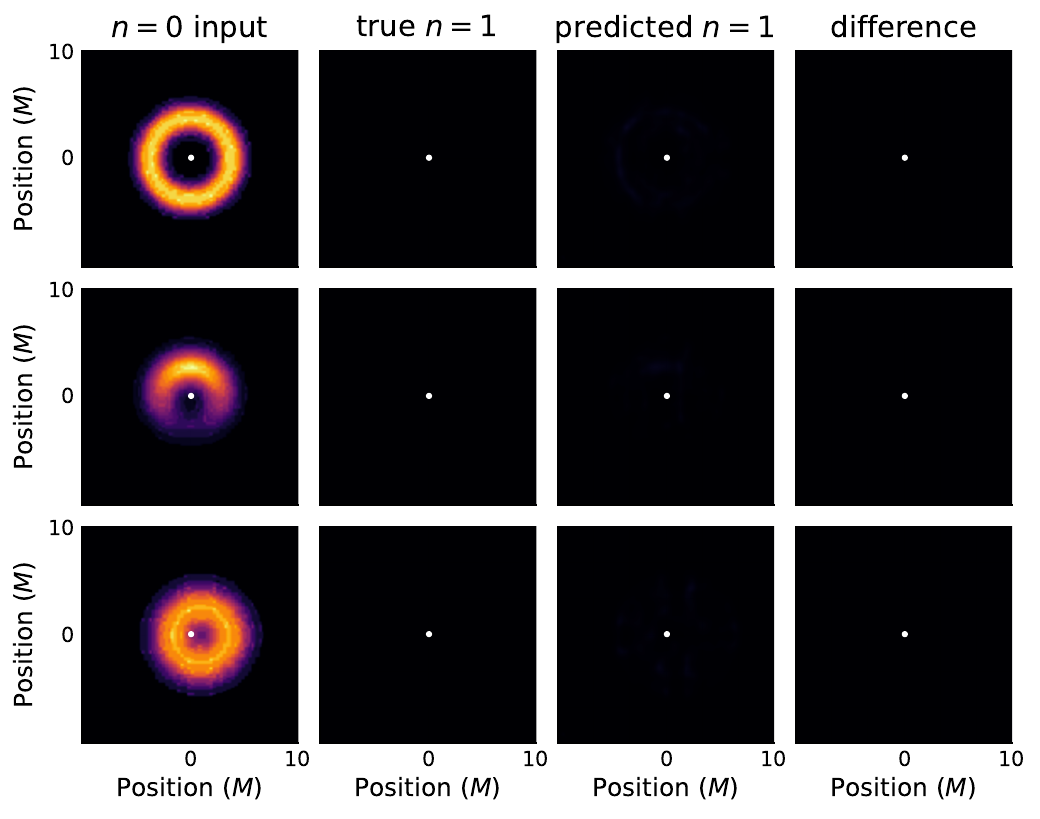}
    \caption{Sample results for inputs without an $n=1$ subring. (Left) input images containing only the $n=0$ subring, (center left) ground truth $n=1$ subrings, (center right) CNN predicted $n=1$ subrings, and (right) corresponding difference images. Each row shows a different parameter configuration: (top, center, bottom) $F_0=$ (0.5, 0.5, 0.5) Jy, $d_0=$ (10.5, 7.9, 7.9) M, $\alpha_0$ = (1.3, 1.6, 2) M, $\Delta x_0$ = (0, 0, 0.7) M, $\beta_{s,0}$ = (0, 0.4, 0); $\theta_0$ = (0, 180, 180)$^\circ$.} The absence of structure in the predictions confirms that the CNN does not falsely generate an $n=1$ subring when it is not present in the input.
    \label{fig:mring_no_n=1_with_diff}
\end{figure*}

An additional but powerful functionality is that the CNN functions as a detector, not predicting an $n=1$ subring when the input image lacks one (Figure \ref{fig:mring_no_n=1_with_diff}). This capability is critical for future BHEX observations as it reduces false positives and ensures spin measurements are derived only from the $n = 1$ subring, not dominant $n=0$ subring emission. The ability to place constraints on the presence of the $n=1$ subring in addition to measurements of its morphology paves the way for possible null hypothesis tests of general relativity.

To test generalization of the CNN, we apply the CNN to black hole images generated by \texttt{KerrBAM} \citep{palumbo2022}, a semi-analytic simulation engine, as well as images produced by general relativistic magnetohydrodynamic (GRMHD) simulations \citep{dexter2012, moscibrodzka2016, moscibrodzka2017, ryan2018, davelaar2018, chael2019b, davelaar2019}. All images are computed at an image resolution of 128x128 pixels, matching the resolution used during CNN training. We focus on magnetically arrested disk (MAD) models \citep{kogan&ruzmaikin1976, narayan2003} and exclude standard and normal evolution (SANE) models \citep{narayan2012} because their pronounced off-axis and non-equatorial emission introduces complexity that would require additional training to model the $n=1$ subring accurately.

\begin{figure*}[ht]
    \centering
    \includegraphics[width=0.7\textwidth]{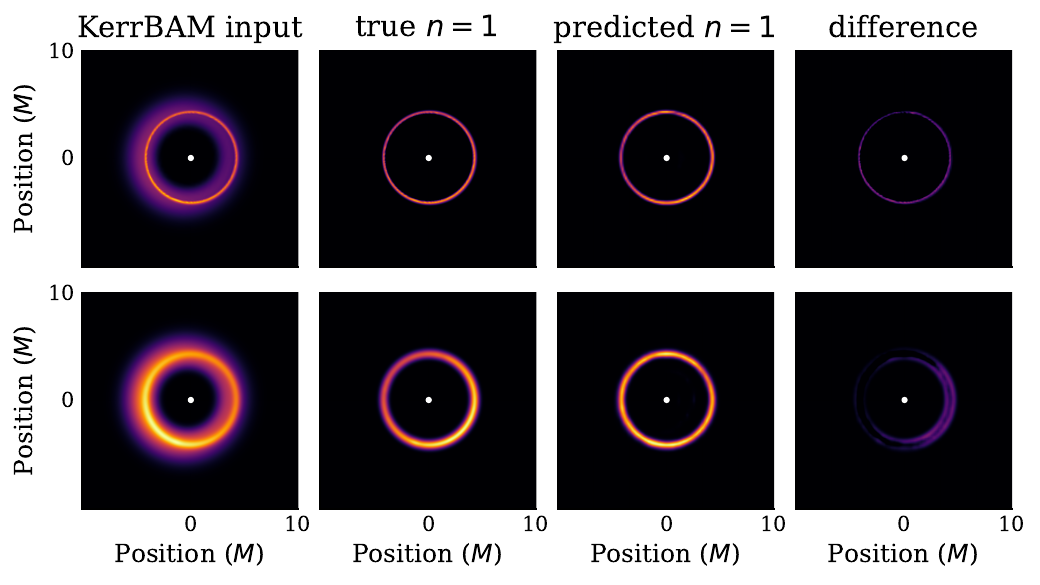}\hfill
    \caption{Black hole images generated by \texttt{KerrBAM} used to evaluate CNN performance. (Left) input images with and without convolution with the nominal BHEX beam, (center left) ground truth $n=1$ subrings, (center right) CNN predicted $n=1$ subrings, and (right) corresponding difference images. The CNN predictions exhibit intensity inconsistencies but qualitatively captures the overall radial profile in the input image.} 
    \label{fig:kerrbam_final}
\end{figure*}

\begin{figure*}[ht]
    \centering
    \includegraphics[width=0.8\textwidth]{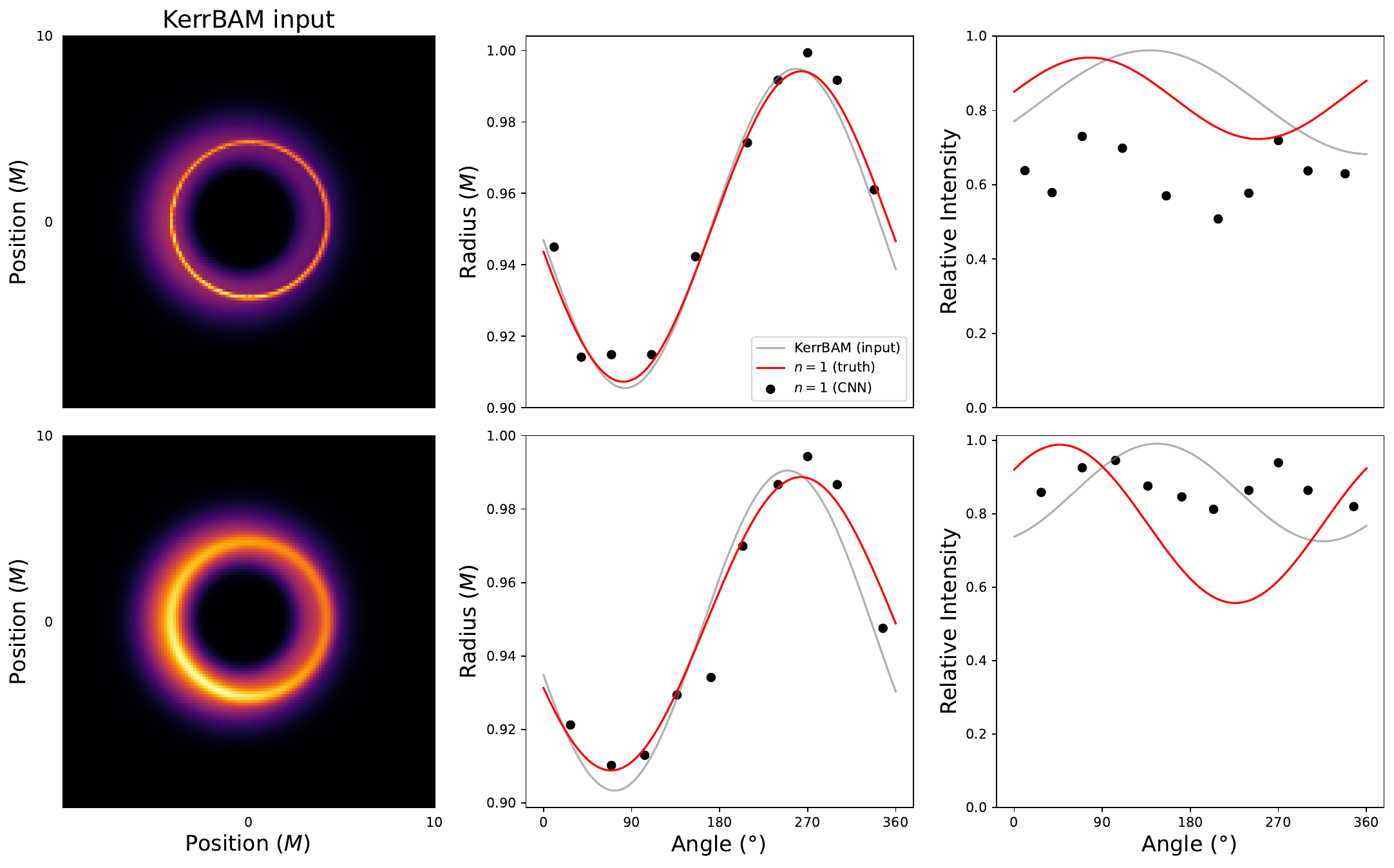}
    \caption{Extraction of radial and angular intensity profiles by \texttt{ringfit} performed on black hole images generated by \texttt{KerrBAM} in Figure \ref{fig:kerrbam_final}. (Left) input images with and without projected BHEX angular resolution; (center) extracted radial profiles for each image's input image, ground truth $n=1$ subring, and CNN predicted $n=1$ subring; and (right) extracted intensity profiles for the same image components. Fits are performed identically as in Figure \ref{fig:feature_extraction}. The CNN accurately recovers the radial profiles of the $n=1$ subring in \texttt{KerrBAM} images of both angular resolutions. However, the CNN fails to accurately recover the brightness profiles of the $n=1$ subring: the brightness extractions for the CNN predicted $n=1$ subring (black) appear biased by the input image extractions (gray), hence do not correspond with the ground truth $n=1$ subring extractions (red).}  
    \label{fig:feature_extraction_kerrbam}
\end{figure*}

Figure \ref{fig:kerrbam_final} shows the CNN’s extraction of the $n=1$ subring from black hole images generated with \texttt{KerrBAM}. Figure \ref{fig:feature_extraction_kerrbam} shows the corresponding feature extraction with \texttt{ringfit}. The black hole images were generated at high spin low inclination ($a=0.8$, $\theta=17^{\circ}$), with the inclination chosen to match that of M87*. We test the CNN on \texttt{KerrBAM} images with and without projected BHEX angular resolution. The predicted $n=1$ subrings for both resolutions overall match the ground truth radial profiles. However, on the images convolved with the BHEX beam, the CNN exhibits difficulties with recovering ground truth intensity profiles as the azimuthal brightness variation of the predicted $n=1$ subring differs significantly from that of the ground truth $n=1$ subring. The large discrepancy between the ground truth and predicted $n=1$ subring intensity profiles under convolution necessitates further training on a wider variety of angular resolutions.


\begin{figure}[h]
    \centering
    \includegraphics[width=0.45\textwidth]{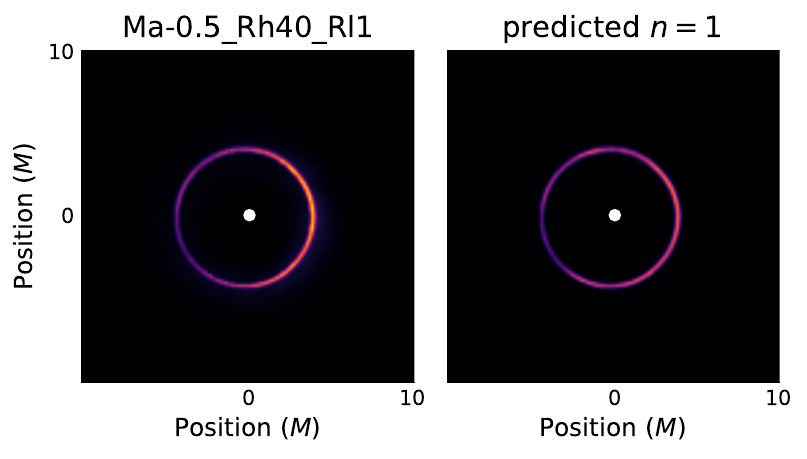}
    \includegraphics[width=0.45\textwidth]{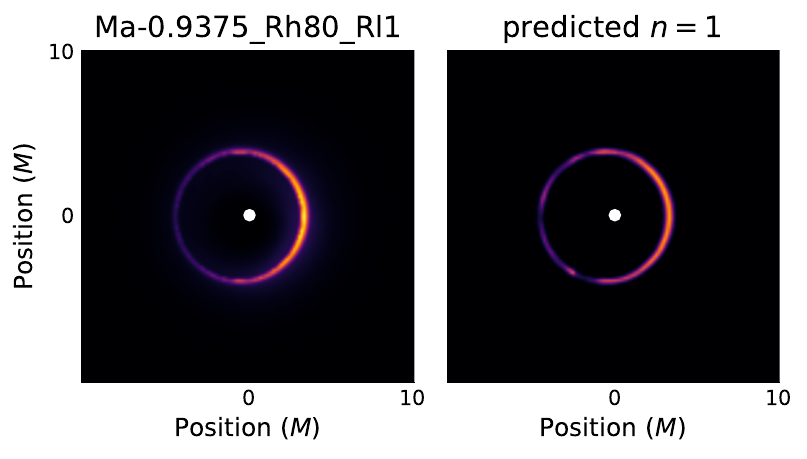}
    \includegraphics[width=0.45\textwidth]{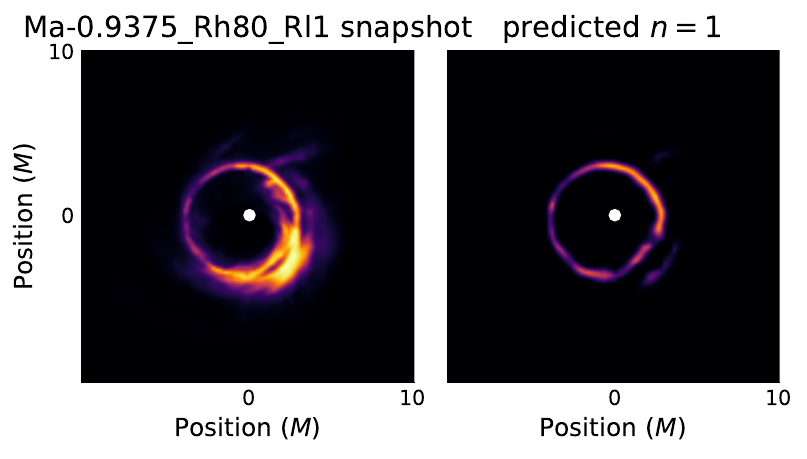}
    
    \caption{Black hole images generated by GRMHD simulations used to evaluate CNN performance. (Top left) time averaged input image simulated with parameters $a=0.5, R_h=40, R_l=1$, (center left) time averaged input image simulated with parameters $a=0.9375, R_h=80, R_l=1$, (bottom left) snapshot input image simulated with parameters $a=0.9375, R_h=80, R_l=1$. For both the time averaged (top right, center right) and snapshot (bottom right) black hole images, the CNN predictions exhibit intensity inconsistencies but qualitatively captures the overall radial profile in the input image, similar to performance on \texttt{KerrBAM} images. However, the CNN prediction for the snapshot image displays larger intensity consistencies and fails to recover fine details.
    %(Left) input images simulated with parameters $a=0.5, R_h=40, R_l=1$ (\textit{top}) and $a=0.9375, R_h=80, R_l=1$ (\textit{center, bottom}). (Right) CNN predicted $n=1$ subrings: like \texttt{KerrBAM} predictions, the CNN predictions exhibit intensity inconsistencies but qualitatively captures the overall radial profile in the input image (\textit{top, center}).
    }
    \label{fig:grmhd tests}
\end{figure}

Figure \ref{fig:grmhd tests} demonstrates the CNN’s extraction of the $n=1$ subring from black hole images generated by GRMHD simulations. The top row corresponds to a black hole with angular momentum $a=0.5$ in natural units, with an electron heating parameter $R_h=40$ in the weakly magnetized disk and $R_l=1$ in the highly magnetized jet \citep{farah2025}. The middle row represents a black hole with $a=0.9375$, $R_h=80$, and $R_l=1$. In both time-averaged images, the CNN cleans much of the diffuse emission, qualitatively capturing the overall radial structure and brightness asymmetry pattern of the $n=1$ subring present in the input images. The bottom row shows the CNN’s extraction of the $n=1$ subring from a single snapshot of a GRMHD simulation, in which additional structures such as transient features are present at short timescales \citep{Johnson_2020}. Larger intensity inconsistencies are visible and fine details are less accurately recovered in the CNN’s prediction, indicating need for training on similar snapshot images. 
%However, the radial profile of the predicted $n=1$ subring appears to match the input image.

While intensity inconsistencies are visible for all tests on KerrBAM and GRMHD simulation images, the predicted $n=1$ subrings display features consistent with expectations for such black hole configurations, highlighting the CNN's robustness and potential applicability to real observational data. These results highlight the CNN’s ability to generalize to unseen simulation images and produce meaningful subring extractions, even when the true $n=1$ subring structure is not known. 

\section{Conclusions} \label{sec:highlight}

The upcoming BHEX mission could present the first opportunity to directly image the black hole photon ring, an important probe of spacetime properties in the vicinity. Efforts to bridge the gap between interferometric measurements of the photon ring and black hole property measurements are actively underway. Recent studies have shown that the $n=1$ photon ring contains sufficient information to make measurements of black hole angular momentum and spin axis orientation. The work presented here provides an important proof-of-concept for a methodology to access that information in an orthogonal way to purely visibility-domain approaches: by removing the $n=0$ emission instead of probing baselines where the $n=1$ subring dominates. 

In this analysis, we presented a neural network architecture which can robustly extract a thin subring from an image contaminated by arbitrary diffuse emission. The CNN isolates the $n=1$ subring from composite input images, preserves physically meaningful features, reliably detects its absence, and generalizes across various simulations. These findings demonstrate the CNN's potential for analysis of high-resolution black hole images, enabling direct spin measurements in future BHEX observations.

In particular, the potential of such a tool is maximized when used in conjunction with other proposed methods. The ability to examine only the narrow $n=1$ subring independent of the $n=0$ contamination can be used with other machine learning tools \citep[e.g.,][]{davelaar2018,farah_2024} to extract spin and inclination in a significantly less complex environment, or with image-domain fits of single-component $m$-rings instead of a more flexible but also more challenging dual-component (narrow and diffuse) fit.

Significant work is needed to improve the utility of the methods beyond the proof-of-concept presented here. In particular, the noise-free zero-convolution training dataset is unrealistic, and additional training on noisy images convolved with the nominal BHEX beam will improve robustness on real data. Additional training data will likely help mitigate the minor failures on some KerrBAM images. The ability to somewhat generalize to complex snapshot images demonstrated here is intriguing but more robust investigation is required to determine the accuracy and potential for integration with long-timescale averaging methods. In total, this framework establishes a foundation that, with continued refinement and realistic training, can evolve into a primary tool for extracting fundamental black hole parameters directly from next-generation horizon-scale images.

% The CNN isolates the $n=1$ subring from composite input images, preserves physically meaningful features, reliably detects its absence, and generalizes across various simulations. These findings demonstrate the CNN's potential for analysis of high-resolution black hole images, enabling direct spin measurements in future BHEX observations.

% Further work to improve the current CNN include training on images with noise and convolution with the BHEX beam to gauge realistic performance and applying the CNN’s capability to KerrBAM images which are generated using real black hole physics.

\begin{acknowledgments}
CD gratefully acknowledges support from a University of California Santa Barbara STEM Edison Academic Research Scholarship. JRF is supported by the National Science Foundation Graduate Research Fellowship Program under Grant No. (NSF 2139319). 
\end{acknowledgments}

\bibliography{references}{}

@article{Farah_2020_BHShadowCurve,
   title={On the Approximation of the Black Hole Shadow with a Simple Polar Curve},
   volume={900},
   ISSN={1538-4357},
   url={http://dx.doi.org/10.3847/1538-4357/aba59a},
   DOI={10.3847/1538-4357/aba59a},
   number={1},
   journal={The Astrophysical Journal},
   publisher={American Astronomical Society},
   author={Farah, Joseph R. and Pesce, Dominic W. and Johnson, Michael D. and Blackburn, Lindy},
   year={2020},
   month=Sept, pages={77} }

@misc{akiyama2026,
      title={Expanding the Horizon of Black Hole Imaging with AtLAST}, 
      author={Kazunori Akiyama and Mariafelicia De Laurentis and Ziri Younsi and Yuto Akiyama and Dominic W. Pesce and Geoffrey C. Bower and Kazuhiro Hada and Jens Kauffmann and Shoko Koyama and Kotaro Moriyama and Derek-Ward Thompson},
      year={2026},
      eprint={2602.04666},
      archivePrefix={arXiv},
      primaryClass={astro-ph.IM},
      url={https://arxiv.org/abs/2602.04666}, 
}

@misc{pesce2019,
      title={Extremely long baseline interferometry with Origins Space Telescope}, 
      author={Dominic W. Pesce and Kari Haworth and Gary J. Melnick and Lindy Blackburn and Maciek Wielgus and Michael D. Johnson and Alexander Raymond and Jonathan Weintroub and Daniel C. M. Palumbo and Sheperd S. Doeleman and David J. James},
      year={2019},
      eprint={1909.01408},
      archivePrefix={arXiv},
      primaryClass={astro-ph.IM},
      url={https://arxiv.org/abs/1909.01408}, 
}

@misc{trippe2026,
      title={The Capella Program: Toward A Space-only High-frequency Radio VLBI Network Formed by Small Satellites in Low Earth Orbits}, 
      author={Sascha Trippe and Taehyun Jung and Jung-Won Lee and Jan Wagner and Jeong-Yeol Han and Doohyon Baek and Wonseok Kang and Jae-Hyun Kyeong and Junghwan Oh and Jae-Young Kim and Jongho Park and Sang-Sung Lee and Jeffrey A. Hodgson and Taeho Kang},
      year={2026},
      eprint={2304.06482},
      archivePrefix={arXiv},
      primaryClass={astro-ph.IM},
      url={https://arxiv.org/abs/2304.06482}, 
}

@inproceedings{Maas2013,
  booktitle={Rectifier Nonlinearities Improve Neural Network Acoustic Models},
  author={Andrew L. Maas},
  year={2013},
  url={https://api.semanticscholar.org/CorpusID:16489696}
}

@misc{loshchilov2019,
      title={Decoupled Weight Decay Regularization}, 
      author={Ilya Loshchilov and Frank Hutter},
      year={2019},
      eprint={1711.05101},
      archivePrefix={arXiv},
      primaryClass={cs.LG},
      url={https://arxiv.org/abs/1711.05101}, 
}

@misc{akiba2019,
      title={Optuna: A Next-generation Hyperparameter Optimization Framework}, 
      author={Takuya Akiba and Shotaro Sano and Toshihiko Yanase and Takeru Ohta and Masanori Koyama},
      year={2019},
      eprint={1907.10902},
      archivePrefix={arXiv},
      primaryClass={cs.LG},
      url={https://arxiv.org/abs/1907.10902}, 
}

@inproceedings{Johnson_2024,
   title={The Black Hole Explorer: motivation and vision},
   url={http://dx.doi.org/10.1117/12.3019835},
   DOI={10.1117/12.3019835},
   booktitle={Space Telescopes and Instrumentation 2024: Optical, Infrared, and Millimeter Wave},
   publisher={SPIE},
   author={Johnson, M. D. and others},
   editor={Coyle, Laura E. and Perrin, Marshall D. and Matsuura, Shuji},
   year={2024},
   month=aug, pages={90} }

@article{Johnson_2020,
   title={Universal interferometric signatures of a black hole’s photon ring},
   volume={6},
   ISSN={2375-2548},
   url={http://dx.doi.org/10.1126/sciadv.aaz1310},
   DOI={10.1126/sciadv.aaz1310},
   number={12},
   journal={Science Advances},
   publisher={American Association for the Advancement of Science (AAAS)},
   author={Johnson, M. D. and others},
   year={2020},
   month=mar }

@misc{farah_2024,
      title={Machine- and deep-learning-driven angular momentum inference from BHEX observations of the $n=1$ photon ring}, 
      author={Farah, J. R. and others},
      year={2024},
      eprint={2411.01060},
      archivePrefix={arXiv},
      primaryClass={astro-ph.HE},
      url={https://arxiv.org/abs/2411.01060}, 
}

@article{lecun_1998,
title = "Gradient-based learning applied to document recognition",
author = "LeCun, Y. and others",
year = "1998",
doi = "10.1109/5.726791",
language = "English (US)",
volume = "86",
pages = "2278--2323",
journal = "Proceedings of the IEEE",
issn = "0018-9219",
publisher = "Institute of Electrical and Electronics Engineers Inc.",
number = "11",
}

@incollection{Krizhevsky_2012,
  author = {Krizhevsky, A. and others},
  booktitle = {Advances in Neural Information Processing Systems 25},
  editor = {Pereira, F. and Burges, C. J. C. and Bottou, L. and Weinberger, K. Q.},
  pages = {1097--1105},
  publisher = {Curran Associates, Inc.},
  title = {ImageNet Classification with Deep Convolutional Neural Networks},
  url = {http://papers.nips.cc/paper/4824-imagenet-classification-with-deep-convolutional-neural-networks.pdf},
  year = 2012
}

@article{Broderick_2022,
   title={Measuring Spin from Relative Photon-ring Sizes},
   volume={927},
   ISSN={1538-4357},
   url={http://dx.doi.org/10.3847/1538-4357/ac4970},
   DOI={10.3847/1538-4357/ac4970},
   number={1},
   journal={The Astrophysical Journal},
   publisher={American Astronomical Society},
   author={Broderick, A. E. and others},
   year={2022},
   month=mar, pages={6} }

@article{EHT_Collab_2022,
   title={First Sagittarius A* Event Horizon Telescope Results. I. The Shadow of the Supermassive Black Hole in the Center of the Milky Way},
   volume={930},
   ISSN={2041-8213},
   url={http://dx.doi.org/10.3847/2041-8213/ac6674},
   DOI={10.3847/2041-8213/ac6674},
   number={2},
   journal={The Astrophysical Journal Letters},
   publisher={American Astronomical Society},
   author={Event Horizon Telescope Collaboration and others},
   year={2022},
   month=may, pages={L12} }

@article{vanderGucht2020,
   title={Deep Horizon: A machine learning network that recovers accreting black hole parameters},
   volume={636},
   ISSN={1432-0746},
   url={http://dx.doi.org/10.1051/0004-6361/201937014},
   DOI={10.1051/0004-6361/201937014},
   journal={Astronomy \& Astrophysics},
   publisher={EDP Sciences},
   author={van der Gucht, J. and others},
   year={2020},
   month=apr, pages={A94} }

@online{chollet2015,
  title={Keras},
  author={Chollet, F. and others},
  year={2015},
  publisher={GitHub},
  url={https://github.com/fchollet/keras},
}

@article{hinton2006,
  title={Reducing the dimensionality of data with neural networks},
  author={Hinton, G. E. and Salakhutdinov, R. R.},
  journal={Science},
  volume={313},
  number={5786},
  pages={504--507},
  year={2006},
  publisher={American Association for the Advancement of Science}
}

@inproceedings{hinton2010,
author = {Nair, Vinod and Hinton, Geoffrey E.},
title = {Rectified linear units improve restricted boltzmann machines},
year = {2010},
isbn = {9781605589077},
publisher = {Omnipress},
address = {Madison, WI, USA},
booktitle = {Proceedings of the 27th International Conference on International Conference on Machine Learning},
pages = {807–814},
numpages = {8},
location = {Haifa, Israel},
series = {ICML'10}
}

@inproceedings{ranzato2007,
  title={Unsupervised Learning of Invariant Feature Hierarchies with Applications to Object Recognition},
  author={Marc'Aurelio Ranzato and Fu Jie Huang and Y-Lan Boureau and Yann LeCun},
  booktitle={2007 IEEE Conference on Computer Vision and Pattern Recognition},
  year={2007},
  pages={1-8}
}

@article{farah2022,
   title={Selective Dynamical Imaging of Interferometric Data},
   volume={930},
   ISSN={2041-8213},
   url={http://dx.doi.org/10.3847/2041-8213/ac6615},
   DOI={10.3847/2041-8213/ac6615},
   number={2},
   journal={The Astrophysical Journal Letters},
   publisher={American Astronomical Society},
   author={Farah, Joseph and Galison, Peter and Akiyama, Kazunori and Bouman, Katherine L. and Bower, Geoffrey C. and Chael, Andrew and Fuentes, Antonio and Gómez, José L. and Honma, Mareki and Johnson, Michael D. and Kofuji, Yutaro and Marrone, Daniel P. and Moriyama, Kotaro and Narayan, Ramesh and Pesce, Dominic W. and Tiede, Paul and Wielgus, Maciek and Zhao, Guang-Yao and Alberdi, Antxon and Alef, Walter and Algaba, Juan Carlos and Anantua, Richard and Asada, Keiichi and Azulay, Rebecca and Baczko, Anne-Kathrin and Ball, David and Baloković, Mislav and Barrett, John and Benson, Bradford A. and Bintley, Dan and Blackburn, Lindy and Blundell, Raymond and Boland, Wilfred and Boyce, Hope and Bremer, Michael and Brinkerink, Christiaan D. and Brissenden, Roger and Britzen, Silke and Broderick, Avery E. and Broguiere, Dominique and Bronzwaer, Thomas and Bustamente, Sandra and Byun, Do-Young and Carlstrom, John E. and Chan, Chi-kwan and Chatterjee, Koushik and Chatterjee, Shami and Chen, Ming-Tang and Chen 陈, Yongjun 永 军 and Cho, Ilje and Christian, Pierre and Conway, John E. and Cordes, James M. and Crawford, Thomas M. and Crew, Geoffrey B. and Cruz-Osorio, Alejandro and Cui, Yuzhu and Davelaar, Jordy and Laurentis, Mariafelicia De and Deane, Roger and Dempsey, Jessica and Desvignes, Gregory and Doeleman, Sheperd S. and Eatough, Ralph P. and Falcke, Heino and Fish, Vincent L. and Fomalont, Ed and Ford, H. Alyson and Fraga-Encinas, Raquel and Friberg, Per and Fromm, Christian M. and Gammie, Charles F. and Garc’a, Roberto and Gentaz, Olivier and Goddi, Ciriaco and Gold, Roman and Gómez-Ruiz, Arturo I. and Gu 顾, Minfeng 敏 峰 and Gurwell, Mark and Hada, Kazuhiro and Haggard, Daryl and Hecht, Michael H. and Hesper, Ronald and Ho 何, Luis C. 子 山 and Ho, Paul and Huang, Chih-Wei L. and Huang 黄, Lei 磊 and Hughes, David H. and Ikeda, Shiro and Inoue, Makoto and Issaoun, Sara and James, David J. and Jannuzi, Buell T. and Janssen, Michael and Jeter, Britton and Jiang 江, Wu 悟 and Jimenez-Rosales, Alejandra and Jorstad, Svetlana and Jung, Taehyun and Karami, Mansour and Karuppusamy, Ramesh and Kawashima, Tomohisa and Keating, Garrett K. and Kettenis, Mark and Kim, Dong-Jin and Kim, Jae-Young and Kim, Jongsoo and Kim, Junhan and Kino, Motoki and Koay, Jun Yi and Koch, Patrick M. and Koyama, Shoko and Kramer, Carsten and Kramer, Michael and Krichbaum, Thomas P. and Kuo, Cheng-Yu and Lauer, Tod R. and Lee, Sang-Sung and Levis, Aviad and Li, Yan-Rong and Li 李, Zhiyuan 志 远 and Lico, Rocco and Lindahl, Greg and Lindqvist, Michael and Liu 刘, Jun 俊 and Liu, Kuo and Liuzzo, Elisabetta and Lo, Wen-Ping and Lobanov, Andrei P. and Loinard, Laurent and Lonsdale, Colin and Lu 路, Ru-Sen 如 森 and MacDonald, Nicholas R. and Mao 毛, Jirong 基 荣 and Marchili, Nicola and Markoff, Sera and Marscher, Alan P. and Martí-Vidal, Iván and Matsushita, Satoki and Matthews, Lynn D. and Medeiros, Lia and Menten, Karl M. and Mizuno, Izumi and Mizuno, Yosuke and Moran, James M. and Moscibrodzka, Monika and Müller, Cornelia and Mejas, Alejandro Mus and Musoke, Gibwa and Nagai, Hiroshi and Nagar, Neil M. and Nakamura, Masanori and Narayanan, Gopal and Natarajan, Iniyan and Nathanail, Antonios and Neilsen, Joey and Neri, Roberto and Ni, Chunchong and Noutsos, Aristeidis and Nowak, Michael A. and Okino, Hiroki and Olivares, Héctor and Ortiz-León, Gisela N. and Oyama, Tomoaki and zel, Feryal and Palumbo, Daniel C. M. and Park, Jongho and Patel, Nimesh and Pen, Ue-Li and Piétu, Vincent and Plambeck, Richard and PopStefanija, Aleksandar and Porth, Oliver and Pötzl, Felix M. and Prather, Ben and Preciado-López, Jorge A. and Psaltis, Dimitrios and Pu, Hung-Yi and Ramakrishnan, Venkatessh and Rao, Ramprasad and Rawlings, Mark G. and Raymond, Alexander W. and Rezzolla, Luciano and Ripperda, Bart and Roelofs, Freek and Rogers, Alan and Ros, Eduardo and Rose, Mel and Roshanineshat, Arash and Rottmann, Helge and Roy, Alan L. and Ruszczyk, Chet and Rygl, Kazi L. J. and Sánchez, Salvador and Sánchez-Arguelles, David and Sasada, Mahito and Savolainen, Tuomas and Schloerb, F. Peter and Schuster, Karl-Friedrich and Shao, Lijing and Shen 沈, Zhiqiang 志 强 and Small, Des and Sohn, Bong Won and SooHoo, Jason and Sun 孙, He 赫 and Tazaki, Fumie and Tetarenko, Alexandra J. and Tilanus, Remo P. J. and Titus, Michael and Toma, Kenji and Torne, Pablo and Traianou, Efthalia and Trent, Tyler and Trippe, Sascha and Bemmel, Ilse van and van Langevelde, Huib Jan and van Rossum, Daniel R. and Wagner, Jan and Ward-Thompson, Derek and Wardle, John and Weintroub, Jonathan and Wex, Norbert and Wharton, Robert and Wiik, Kaj and Wong, George N. and Wu, Qingwen and Yoon, Doosoo and Young, André and Young, Ken and Younsi, Ziri and Yuan 袁, Feng 峰 and Yuan 袁, Ye-Fei 业 飞 and Zensus, J. Anton and Zhao, Shan-Shan},
   year={2022},
   month=may, pages={L18} }

@article{eht2019,
   title={First M87 Event Horizon Telescope Results. IV. Imaging the Central Supermassive Black Hole},
   volume={875},
   ISSN={2041-8213},
   url={http://dx.doi.org/10.3847/2041-8213/ab0e85},
   DOI={10.3847/2041-8213/ab0e85},
   number={1},
   journal={The Astrophysical Journal Letters},
   publisher={American Astronomical Society},
   author={Event Horizon Telescope Collaboration and others},
   year={2019},
   month=apr, pages={L4} }

@software{eht-imaging,
  author       = {Chael, Andrew},
  title        = {eht-imaging},
  month        = jan,
  year         = 2025,
  publisher    = {Zenodo},
  version      = {v1.2.10},
  doi          = {10.5281/zenodo.14624987},
  url          = {https://doi.org/10.5281/zenodo.14624987},
  swhid        = {swh:1:dir:11b0a450b955521bcf69c0dc88db464305a8f18e
                   ;origin=https://doi.org/10.5281/zenodo.2614008;vis
                   it=swh:1:snp:3be039079f73c6cd8d8314789a2a95f284e0f
                   f03;anchor=swh:1:rel:c80c8f29e8e4b7c6cacfe7619fff0
                   cec1ed182a6;path=achael-eht-imaging-c38f16f
                  },
}

@article{palumbo2022,
   title={Photon Ring Symmetries in Simulated Linear Polarization Images of Messier 87*},
   volume={929},
   ISSN={1538-4357},
   url={http://dx.doi.org/10.3847/1538-4357/ac59b4},
   DOI={10.3847/1538-4357/ac59b4},
   number={1},
   journal={The Astrophysical Journal},
   publisher={American Astronomical Society},
   author={Palumbo, Daniel C. M. and Wong, George N.},
   year={2022},
   month=apr, pages={49} }

@ARTICLE{kogan&ruzmaikin1976,
       author = {{Bisnovatyi-Kogan}, G.~S. and {Ruzmaikin}, A.~A.},
        title = "{The Accretion of Matter by a Collapsing Star in the Presence of a Magnetic Field. II: Self-consistent Stationary Picture}",
      journal = {\apss},
         year = 1976,
        month = jul,
       volume = {42},
       number = {2},
        pages = {401-424},
          doi = {10.1007/BF01225967},
       adsurl = {https://ui.adsabs.harvard.edu/abs/1976Ap&SS..42..401B}
}

@ARTICLE{narayan2003,
       author = {{Narayan}, Ramesh and {Igumenshchev}, Igor V. and {Abramowicz}, Marek A.},
        title = "{Magnetically Arrested Disk: an Energetically Efficient Accretion Flow}",
      journal = {\pasj},
         year = 2003,
        month = dec,
       volume = {55},
        pages = {L69-L72},
          doi = {10.1093/pasj/55.6.L69},
archivePrefix = {arXiv},
       eprint = {astro-ph/0305029},
 primaryClass = {astro-ph},
       adsurl = {https://ui.adsabs.harvard.edu/abs/2003PASJ...55L..69N}
}

@ARTICLE{narayan2012,
       author = {{Narayan}, Ramesh and {Sądowski}, Aleksander and {Penna}, Robert F. and {Kulkarni}, Akshay K.},
        title = "{GRMHD simulations of magnetized advection-dominated accretion on a non-spinning black hole: role of outflows}",
      journal = {\mnras},
         year = 2012,
        month = nov,
       volume = {426},
       number = {4},
        pages = {3241-3259},
          doi = {10.1111/j.1365-2966.2012.22002.x},
archivePrefix = {arXiv},
       eprint = {1206.1213},
 primaryClass = {astro-ph.HE},
       adsurl = {https://ui.adsabs.harvard.edu/abs/2012MNRAS.426.3241N}
}

@article{dexter2012,
author = {Dexter, Jason and McKinney, Jonathan C. and Agol, Eric},
title = {The size of the jet launching region in M87},
journal = {Monthly Notices of the Royal Astronomical Society},
volume = {421},
number = {2},
pages = {1517-1528},
doi = {https://doi.org/10.1111/j.1365-2966.2012.20409.x},
url = {https://onlinelibrary.wiley.com/doi/abs/10.1111/j.1365-2966.2012.20409.x},
eprint = {https://onlinelibrary.wiley.com/doi/pdf/10.1111/j.1365-2966.2012.20409.x},
year = {2012}
}

@article{moscibrodzka2016,
    author = "Moscibrodzka, Monika and Falcke, Heino and Shiokawa, Hotaka",
    title = "{General relativistic magnetohydrodynamical simulations of the jet in M 87}",
    eprint = "1510.07243",
    archivePrefix = "arXiv",
    primaryClass = "astro-ph.HE",
    doi = "10.1051/0004-6361/201526630",
    journal = "Astron. Astrophys.",
    volume = "586",
    pages = "A38",
    year = "2016"
}

@article{moscibrodzka2017,
author = {Moscibrodzka, Monika and Dexter, Jason and Davelaar, Jordy and Falcke, Heino},
year = {2017},
month = {03},
pages = {},
title = {Faraday rotation in GRMHD simulations of the jet launching zone of M87},
volume = {468},
journal = {Monthly Notices of the Royal Astronomical Society},
doi = {10.1093/mnras/stx587}
}

@article{ryan2018,
   title={Two-temperature GRRMHD Simulations of M87},
   volume={864},
   ISSN={1538-4357},
   url={http://dx.doi.org/10.3847/1538-4357/aad73a},
   DOI={10.3847/1538-4357/aad73a},
   number={2},
   journal={The Astrophysical Journal},
   publisher={American Astronomical Society},
   author={Ryan, Benjamin R. and Ressler, Sean M. and Dolence, Joshua C. and Gammie, Charles and Quataert, Eliot},
   year={2018},
   month=sep, pages={126} }

@article{davelaar2018,
   title={General relativistic magnetohydrodynamical κ-jet
models for Sagittarius A*},
   volume={612},
   ISSN={1432-0746},
   url={http://dx.doi.org/10.1051/0004-6361/201732025},
   DOI={10.1051/0004-6361/201732025},
   journal={Astronomy \& Astrophysics},
   publisher={EDP Sciences},
   author={Davelaar, J. and Mościbrodzka, M. and Bronzwaer, T. and Falcke, H.},
   year={2018},
   month=apr, pages={A34} }

@software{chael2019b,
       author = {{Chael}, Andrew A. and {Bouman}, Katherine L. and {Johnson}, Michael D. and {Narayan}, Ramesh and {Doeleman}, Sheperd S. and {Wardle}, John F.~C. and {Blackburn}, Lindy L. and {Akiyama}, Kazunori and {Wielgus}, Maciek and {Chan}, Chi-kwan and {Farah}, Joseph R. and {Palumbo}, Daniel and {Pesce}, Dominic},
        title = "{ehtim: Imaging, analysis, and simulation software for radio interferometry}",
 howpublished = {Astrophysics Source Code Library, record ascl:1904.004},
         year = 2019,
        month = apr,
          eid = {ascl:1904.004},
archivePrefix = {ascl},
       eprint = {1904.004},
       adsurl = {https://ui.adsabs.harvard.edu/abs/2019ascl.soft04004C}
}

@article{davelaar2019,
  title={Modeling non-thermal emission from the jet-launching region of M 87 with adaptive mesh refinement},
  author={Davelaar, Jordy and Olivares, Hector and Porth, Oliver and Bronzwaer, Thomas and Janssen, Michael and Roelofs, Freek and Mizuno, Yosuke and Fromm, Christian M and Falcke, Heino and Rezzolla, Luciano},
  journal={Astronomy \& Astrophysics},
  volume={632},
  pages={A2},
  year={2019},
  publisher={EDP Sciences}
}

@misc{chang2024,
      title={Bayesian Black Hole Photogrammetry}, 
      author={Dominic O. Chang and Michael D. Johnson and Paul Tiede and Daniel C. M. Palumbo},
      year={2024},
      eprint={2405.04749},
      archivePrefix={arXiv},
      primaryClass={astro-ph.HE},
      url={https://arxiv.org/abs/2405.04749}, 
}

@article{keeble2025,
  title = {Inferring black hole spin from interferometric measurements of the first photon ring: A geometric approach},
  author = {Keeble, Lennox S. and C\'ardenas-Avenda\~no, Alejandro and Palumbo, Daniel C. M.},
  journal = {Phys. Rev. D},
  volume = {111},
  issue = {10},
  pages = {103042},
  numpages = {19},
  year = {2025},
  month = {May},
  publisher = {American Physical Society},
  doi = {10.1103/PhysRevD.111.103042},
  url = {https://link.aps.org/doi/10.1103/PhysRevD.111.103042}
}

@article{Paugnat_2022,
   title={Photon ring test of the Kerr hypothesis: Variation in the ring shape},
   volume={668},
   ISSN={1432-0746},
   url={http://dx.doi.org/10.1051/0004-6361/202244216},
   DOI={10.1051/0004-6361/202244216},
   journal={Astronomy \& Astrophysics},
   publisher={EDP Sciences},
   author={Paugnat, H. and Lupsasca, A. and Vincent, F. H. and Wielgus, M.},
   year={2022},
   month=nov, pages={A11} }

@misc{farah2025,
      title={Interferometric inference of black hole spin from photon ring size and brightness}, 
      author={Joseph R. Farah and Alexandru Lupsasca and Eliot Quataert and Michael D. Johnson},
      year={2025},
      eprint={2509.23628},
      archivePrefix={arXiv},
      primaryClass={astro-ph.HE},
      url={https://arxiv.org/abs/2509.23628}, 
}

@INPROCEEDINGS{Akiyama2024BHEXJapan,
       author = {{Akiyama}, Kazunori and {Niinuma}, Kotaro and {Hada}, Kazuhiro and {Doi}, Akihiro and {Hagiwara}, Yoshiaki and {Higuchi}, Aya E. and {Honma}, Mareki and {Kawashima}, Tomohisa and {Kolev}, Dimitar and {Koyama}, Shoko and {Masui}, Sho and {Ohsuga}, Ken and {Sano}, Hidetoshi and {Takami}, Hideki and {Tsunetoe}, Yuh and {Uzawa}, Yoshinori and {Akahori}, Takuya and {Akiyama}, Yuto and {Galison}, Peter and {Hayashi}, Takayuki J. and {Hirota}, Tomoya and {Inoue}, Makoto and {Iwata}, Yuhei and {Johnson}, Michael D. and {Kino}, Motoki and {Kofuji}, Yutaro and {Mizuno}, Yosuke and {Moriyama}, Kotaro and {Nagai}, Hiroshi and {Nakamura}, Kenta and {Notsu}, Shota and {Ono}, Fumie and {Oya}, Yoko and {Oyama}, Tomoaki and {Rana}, Hannah and {Saida}, Hiromi and {Saito}, Ryo and {Saito}, Yoshihiko and {Sasada}, Mahito and {Sawada-Satoh}, Satoko and {Takahashi}, Mikiya M. and {Takamura}, Mieko and {Tong}, Edward and {Tsuji}, Hiroyuki and {Yoshioka}, Shogo and {Watanabe}, Yoshimasa},
        title = "{The Japanese vision for the Black Hole Explorer mission}",
    booktitle = {Space Telescopes and Instrumentation 2024: Optical, Infrared, and Millimeter Wave},
         year = 2024,
       editor = {{Coyle}, Laura E. and {Matsuura}, Shuji and {Perrin}, Marshall D.},
       series = {Society of Photo-Optical Instrumentation Engineers (SPIE) Conference Series},
       volume = {13092},
        month = aug,
          eid = {130922E},
        pages = {130922E},
          doi = {10.1117/12.3019968},
archivePrefix = {arXiv},
       eprint = {2406.09516},
 primaryClass = {astro-ph.IM},
       adsurl = {https://ui.adsabs.harvard.edu/abs/2024SPIE13092E..2EA}
}

@ARTICLE{Luminet1979,
       author = {{Luminet}, J. -P.},
        title = "{Image of a spherical black hole with thin accretion disk.}",
      journal = {\aap},
         year = 1979,
        month = may,
       volume = {75},
        pages = {228-235},
       adsurl = {https://ui.adsabs.harvard.edu/abs/1979A&A....75..228L}
}

@ARTICLE{GrallaLupsascaLensing,
       author = {{Gralla}, Samuel E. and {Lupsasca}, Alexandru},
        title = "{Lensing by Kerr black holes}",
      journal = {\prd},
         year = 2020,
        month = feb,
       volume = {101},
       number = {4},
          eid = {044031},
        pages = {044031},
          doi = {10.1103/PhysRevD.101.044031},
archivePrefix = {arXiv},
       eprint = {1910.12873},
 primaryClass = {gr-qc},
       adsurl = {https://ui.adsabs.harvard.edu/abs/2020PhRvD.101d4031G}
}

@software{ringfit,
  author       = {Myhre, Frank and Farah, Joseph},
  title        = {ringfit},
  year         = 2025,
  version      = {v0.4.5},
}
\bibliographystyle{aasjournalv7}

\end{document}